\documentclass[11pt,a4paper]{article}
\usepackage[utf8]{inputenc}
\usepackage{graphicx}
\usepackage{authblk}
\usepackage[version=3]{mhchem}
\usepackage[left=3cm,right=3cm,top=3cm,bottom=3cm]{geometry}

\begin{document}
\title{Thin polymerized C$_{60}$ coatings deposited in electrostatic field via electron-beam dispersion of fullerite}
\author[1,2]{Ihar Razanau \thanks {igor.ryazanov@live.jp}}
\author[1]{Tetsu Mieno}
\author[3]{Victor Kazachenko}
\affil[1]{Department of Physics, Shizuoka University, Shizuoka, 422-8529, Japan}
\affil[2]{Francisk Skorina Gomel State University, Gomel, 246019 Belarus}
\affil[3]{Belarusian State University of Transport, Gomel, 246653, Belarus}
\renewcommand\Authands{ and }
\date{15 March 2010}
\maketitle
\begin{abstract}
		The aim of the present study is to clarify the charge composition of fullerite  \ce{C60} electron-beam dispersion (EBD) products and investigate the influence of fullerene ions and electrons on the structure of the deposited coatings by applying an additional electrostatic field to the substrates. It was found that  \ce{C60} EBD products contain positive fullerene ions and electrons. By using Raman and attenuated total reflection Fourier transform infrared spectroscopy, X-ray photoelectron spectroscopy, laser desorption/ionization mass-spectrometry and atomic force microscopy, it was shown that the assistance of the electrons additionally accelerated up to 300~eV results in the formation of a mixture of dumb-bell- and peanut-shaped  \ce{C60} polymers. The assistance of the positive fullerene ions additionally accelerated up to 300~eV leads to the formation of highly cross-linked random 3D networks of covalently bonded fullerene molecules.\\
		
\end{abstract}

\begin{small}
\begin{center}
Published in  Thin Solid Films 519 (2010) 1285--1292\\
DOI: 10.1016/j.tsf.2010.09.027\\
\end{center}
\end{small}

\section{Introduction}
Since the discovery of \ce{C60} photopolymerization in 1993~\cite{1} numerous techniques and methods inducing fullerite structural modification have been found. The irradiation of fullerite by visible or ultraviolet light of sufficient intensity in the absence of oxygen leads to covalent bonding between fullerene molecules through [2+2] cycloaddition reaction (so-called dumb-bell-shaped polymers). The reaction occurs between an excited molecule in the triplet state and a molecule in the ground state. Oxygen is an effective triplet state quencher, and thus, in the presence of atmospheric oxygen, the reaction is negligible~\cite{2}. Up to 6 four-membered ring linkages per \ce{C60} molecule are formed with long irradiation time (more than 200 h), inducing the formation of a saturated 2D-rhombohedral structure~\cite{3}. It has been found that simultaneous deposition and UV light irradiation results in effective photopolymerization of \ce{C60} films up to the degree of photopolymerization of about 90--95\%. Analysis of the vibrational spectroscopy data has revealed the presence of \ce{C60} dimers, linear chains and 2D polymers in the photopolymerized films with dimers and linear polymer chains being the main components~\cite{4, 5}. High-pressure high-temperature (HPHT) treatment of \ce{C60} powder by increasing the interaction among molecules also induces their [2+2] cross-linking~\cite{6, 7}. Depending on the experimental conditions, dimer~(D), linear chain orthorhombic~(O), two-dimensional tetragonal~(T) or two-dimensional rhombohedral~(R) phases are formed~\cite{8, 9, 10}. Superhard and ultrahard 3D carbon networks have been obtained at temperatures of 620--1830 K and pressures of 9.5 and 13 GPa~\cite{11}. Fullerene molecules were still unbroken under these conditions, and no characteristic spectral features of diamond were observed in the obtained material. The polymerization by alkali metal doping~\cite{12} and mechanochemical reaction with the catalyst (\ce{KCN}, \ce{K2CO3}, \ce{CH3CO2K}, Li, Na, K, Mg, Al, Zn) by the use of high speed vibration milling~\cite{13} has been reported as well. It has been found that in metastable \ce{RbC60} and \ce{KC60} compounds, owing to the charge transfer from the alkali metal, dimer anions are formed through a single C–C interfullerene bond~\cite{14}. Zhao \textit{et al.}  observed the electron-stimulated modification of \ce{C60} films induced by both 3~eV electrons from a scanning tunneling microscope tip and 1500~eV electrons from an electron gun~\cite{15}. Tada and Kanayama reported a decrease in the dissolution rate of \ce{C60} films in organic solvents after 20~keV electron-beam irradiation up to the irradiation dose of 0.024 C/cm$^2$~\cite{16}. Hara \textit{et al.} showed that the irradiation of a thin \ce{C60} film with a 3~keV electron beam for several hours leads to the formation of the quasi-one-dimensional “peanut- shaped” \ce{C60} polymers from the initial “dumb-bell” configuration with four-membered carbon ring linkages between the molecules through a series of generalized Stone–Wales (GSW) rearrangements of carbon bonds~\cite{17}. It was found that the peanut-shaped fullerene polymers possess the metallic electronic structure~\cite{18}. Dmytrenko \textit{et al.} reported the degradation of \ce{C60} molecules in thin films subjected to the high-energy electron irradiation ($E=1.8$ MeV) up to the irradiation dose of 8 MGy~\cite{19}. \\

We have previously shown~\cite{20} that thin fullerene polymer films are formed by electron-beam dispersion (EBD) of fullerite powder in vacuum. This method combines film deposition with film polymerization in one process. The main feature of the method is that \ce{} layers grow from the active gas phase containing various active species (excited molecules, ions and electrons). The presence of charged particles in the molecular flow and their crucial role in material modification processes allow us to control polymerization reactions by applying additional electric field to the substrates on which  \ce{C60} films are deposited. However, neither the charge composition of  \ce{C60} EBD products nor the difference between the role of fullerene ions and electrons in the modification of the film structure during its growth from the active gas phase have been studied. The aim of the present study is to clarify the charge composition of fullerite EBD products and the role of fullerene ions and electrons in fullerene polymer film formation.

\section{Experimental details}
Thin fullerene films (about 200~nm) were deposited from the active gas phase produced by the EBD of fullerite \ce{C60} in the experimental vacuum setup shown in Fig. 1. Coatings were deposited in a stainless-steel vacuum chamber 208~mm in diameter and 240~mm in height evacuated using a turbomolecular pump to a pressure of $1\cdot10^{-2}$~Pa. For each deposition experiment \ce{C60} powder (purity $>99.5$ wt.\%) was pressed into a disc target  5~mm in diameter to avoid fullerite scattering due to charging under electron beam irradiation and to maintain the constant size of the evaporating surface. The applied pressure was low (25~MPa) and insufficient for considerable fullerite material molecular structure modification. Before the deposition process, the fullerite target was degassed for 1 hour in the vacuum chamber by maintaining the heating table temperature of 200 $^\circ$C. An electron beam with the energy of 1.2~keV, produced by an electron gun, enabled EBD of the fullerite target. Uniform irradiation of the target was achieved owing to the size of the electron beam spot being larger than the size of the fullerite target. A quartz crystal microbalance (QCM) was used to control the film thickness during its growth from the volatile products of EBD. All studied coatings were deposited at the growth rate of about 1--2.5~nm/min.\\

\begin{figure}[h!]
 \begin{center}
 		\includegraphics[scale=1]{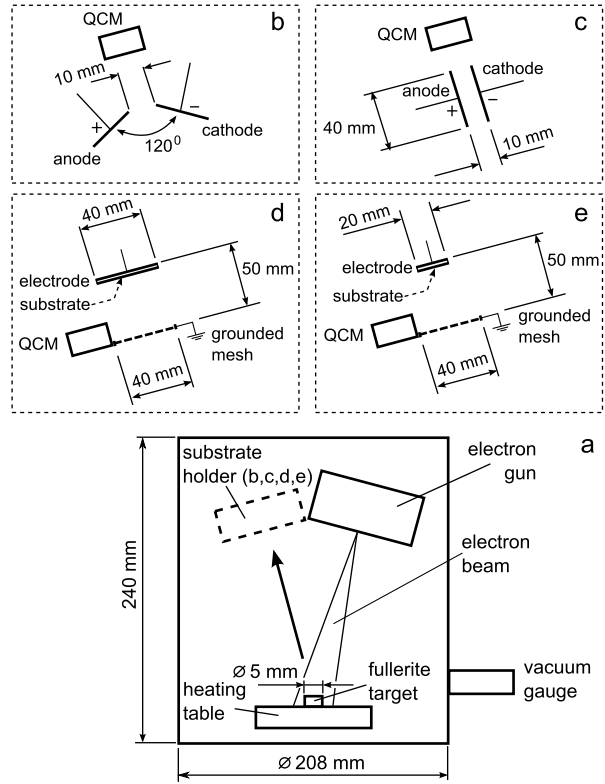}
		\caption{Schematic representation of experimental setup for fullerene film deposition by electron-beam dispersion of the fullerite target (a) and 4 types of substrate holders used in the experiment (b, c, d, e).}
\end{center} 
\end{figure}

A constant electric field was applied to the substrate during film deposition. Four different substrate holder schemes were used in the experiment. These deposition schemes allowed us to determine the type of charged particles in the fullerite EBD products, and estimate their energy and their influence on the structure of deposited layers. In scheme 1 (Fig. 1b), two electrodes are at the angle of about 1200 to each other to spread the electrostatic field widely into the area of evaporating EBD products. In scheme 2 (Fig. 1c) electrodes are parallel to each other at the distance of 10~mm. The size of the electrodes is 4 times larger (40x40~mm) than the distance between them and therefore, in this scheme, the electrostatic field is mostly concentrated between the electrodes. In both schemes, fullerene films were formed directly on the copper electrodes. The voltage of 300~V was applied between the electrodes during deposition and the QCM sensor was placed behind the electrodes.\\

In schemes 3 and 4 (Figs. 1d,e), a fullerene coating is deposited onto the substrates mounted on the electrode placed behind a grounded copper~mesh (0.8~mm diameter, 2~mesh/cm) at the distance of 50~mm. Aluminum film (for Raman spectroscopy and mass spectrometry), aluminum-coated poly-(ethyleneterephthalate) (PET) film (for attenuated total reflection FTIR spectroscopy) and a Si monocrystal with a thermally deposited 100-nm-thick Al sublayer (for atomic force microscopy) were used as the substrates. Constant electric potentials of +10, 0, -10, -50, -100, -200 and -300~V were applied to the electrode during coating deposition. In scheme 3, the electrode has the same area as the grounded~mesh (40x40~mm). The scheme enables the acceleration/deceleration of charged particles depending on the applied potential and the particle charge sign. In scheme 4, the electrode has 4 times a smaller area (20x20~mm) in comparison with the grounded~mesh. Because of electrostatic focusing, this scheme enables not just the acceleration/deceleration of charged particles but an increase of their flux as well. In schemes 3 and 4, the QCM sensor is placed at the level of the grounded~mesh.\\

Raman spectroscopy measurement was performed using a Jasco NR1800 spectrometer equipped with a cooled CCD detector in a backscattering geometry. The 488~nm line of an argon ion laser was used as the excitation source. As mentioned before, the irradiation of fullerite by UV-Vis light can induce polymerization of the material. To avoid laser-induced coating modification, a test measurement on a nonpolymerized \ce{C60} film was carried out. No shift of the A$_g$(2) Raman line at 1468~cm$^{-1}$ (an intrinsic pentagonal pinch mode of fullerene monomers) was detected upon laser irradiation. Thus the laser at the power level used induces negligible film polymerization during the measurement. Attenuated total reflection (ATR) FTIR measurement was performed using a Bruker Vertex 70 spectrometer with a Carl Zeiss ATR unit and a KRS-5 crystal. An ESCA 3400 apparatus with a Mg K$\alpha$ (1253.6~eV) X-ray source was used for X-ray photoelectron spectroscopy (XPS) measurement. The base pressure in the XPS system was kept at $1\cdot10^{-6}$~Pa during the measurement. Laser desorption/ionization (LDI) mass-spectrometry measurement was performed using a Bruker AutoFlex time-of-flight mass spectrometer. Atomic force microscopy (AFM) measurement was performed using an SPI 3800N probe station (Seiko Instruments Inc.). The dynamic force mode was applied for the measurement. This mode allows us to obtain surface topography data and a map of the surface mechanical properties (phase contrast images) simultaneously.
\section{Results and discussion}

The voltage of 300~V applied between the electrodes in scheme 1 during fullerite EBD allows cathode and anode coatings to be formed with the assistance of additionally accelerated positive and negative charged particles, respectively. A cathode layer is thought to grow under a bombardment of positive fullerene ions, and the anode layer under a bombardment of negative fullerene ions and electrons. Both cathode and anode coatings after deposition were insoluble in toluene, indicating significant molecular structure modification in comparison with pristine \ce{C60}. Unlike the anode layer, the cathode coating possessed a characteristic black color. The Raman spectra of both cathode and anode coatings (Fig. 2a) significantly differ from the spectrum of pristine fullerite \ce{C60}.\\

The Raman active A$_g$(2) pentagonal pinch mode of \ce{C60} monomers at 1468~cm$^{-1}$ is a good indicator of polymerization reactions. The formation of covalent bonds between \ce{C60} molecules dramatically reduces molecular symmetry, giving rise to additional bands, the shift of active modes, the activation of silent modes, and the splitting of the degenerate modes. In the HPHT experiments, it was shown that upon polymerization, the strongest Raman A$_g$(2) line shifts towards lower frequencies to 1462~cm$^{-1}$ for dimers, 1457~cm$^{-1}$ for linear chains, and 1449 and 1406~cm$^{-1}$ for tetragonal and rhombohedral 2D polymer networks~\cite{10}, and therefore this spectral region is distinct for each kind of polymer. However, in the case of a sample containing different kinds of polymers the spectrum in the vicinity of the original A$_g$(2) line consists of several modes that are poorly resolved at room temperature, and hence, a more detailed spectrum analysis is necessary.\\

Lorentzian lineshape fitting (Fig. 2b) of the Raman spectrum of the anode coating shows a number of bands with frequencies in good agreement with the Raman data available for HPHT polymers. The spectrum contains Raman bands characteristic of fullerene monomers, dimers, linear chains and 2D polymer networks. Such polymerized fullerite layers containing different types of \ce{C60} polymers are similar to the layers obtained previously by us without additional electric field assistance~\cite{20}.\\

\begin{figure}[h!]
 \begin{center}
 		\includegraphics[scale=1]{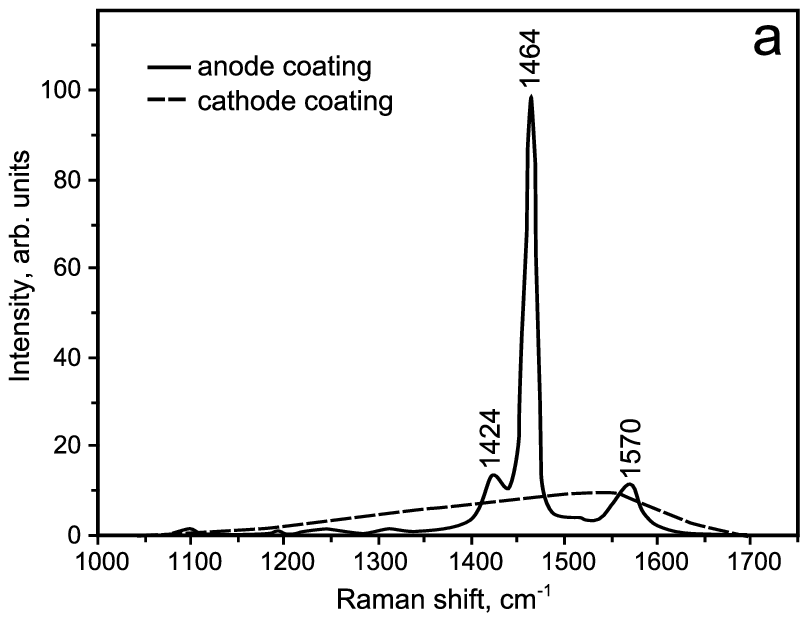}
 		\includegraphics[scale=1]{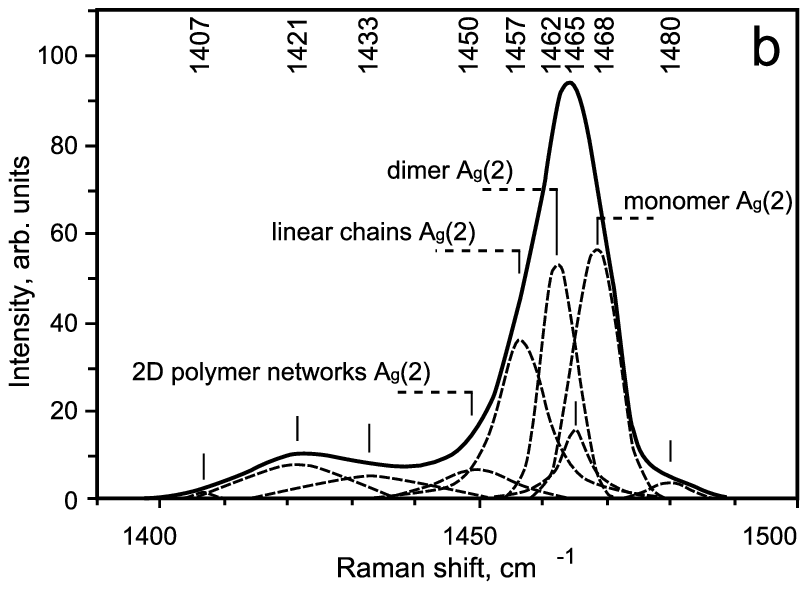}
		\caption{Raman spectra of the coatings deposited using scheme 1 onto the anode and the cathode (a), Lorentzian line-shape analysis of the anode coating Raman spectrum in the range of 1400--1500~cm$^{-1}$ (b).}
\end{center} 
\end{figure}

Unlike the anode coating, the Raman spectrum of the cathode coating exhibits a broad continuum in the range of 1000--1700~cm$^{-1}$ (Fig. 2a). It should be noted that Raman spectra of hydrogen-free amorphous carbon coatings also show a broad band consisting of so-called D and G peaks in this spectral region~\cite{21}. However, the Lorentzian lineshape fitting of the region with two components at around 1350~cm$^{-1}$ and 1510--1600~cm$^{-1}$ for D and G modes, respectively, gives only poor agreement with the experimental spectrum. On the other hand, the described Raman spectrum strongly resembles spectra of the superhard carbon phases produced by Blank \textit{et al.} using HPHT treatment, particularly the sample obtained under the pressure of 9.5 GPa and the temperature of 670 K~\cite{11}. The material was characterized by the authors as a random network of covalently linked fullerene molecules. We will discuss later whether or not the fullerene molecules are still characteristic structural elements of the described cathode coating.\\

Significant structure modification of the cathode layer and conventional polymerization of the anode layer clearly indicate that the charged phase of fullerite EBD products contain positive fullerene ions and electrons. Indeed, heavy positive fullerene ions additionally accelerated to the energies of 300~eV by the electrostatic field bombard the growing cathode layer, leading to the formation of the insoluble strongly cross-linked carbon coating. On the other hand, additional acceleration of the electrons insignificantly affects polymerization reactions induced by them in the growing anode layer.\\

In scheme 2, the electrodes form a planar capacitor with the electric field concentrated between them, and charged particles are attracted and accelerated by the corresponding electrode only in the interelectrode space. Considering that the 10~mm gap between the cathode and the anode is small in comparison with the 120~mm distance to the target (5~mm in diameter), we can assume the part of molecular flow entering the capacitor to be parallel to the electrodes. Under this assumption, the original kinetic energy of charged particles can be estimated from the distance that they travel in the direction parallel to the electrode surface before being captured by it. The applied voltage of 300~V allows \ce{C60+} ions and electrons with energies of up to 1200~eV to be separated from the molecular flow. The coating growth rate registered by the QCM sensor placed behind the electrodes was decreasing by 1-3\% on average upon applying the voltage and returning to its original value after turning off the voltage. This observation proves that fullerite EBD products contain ions and shows that the ion content under the EBD conditions used is a few mass percent.\\

The coating deposited onto the anode using scheme 2 had a uniform color, whereas the cathode coating exhibited a contrasting black stripe near the front edge of the electrode. Raman spectra taken at different distances from the front edge of the cathode show the gradual transformation from the broad continuum at 1100--1700~cm$^{-1}$ with weak fullerene features at around 1424 and 1464~cm$^{-1}$ to the spectrum dominated by these fullerene features (Fig. 3). Apparently, the gradient of the coating structure is caused by the gradual decrease of the dose of positive ion bombardment, owing to a specific distribution of their original kinetic energy. Indeed, the higher the ion energy, the longer the distance it travels in the interelectrode space. Ions with the same energies bombard the cathode surface approximately at the same distance from its front edge causing modification of the growing layer. The higher the ion bombardment dose, the stronger the modification of the coating.\\

\begin{figure}[h!]
 \begin{center}
 		\includegraphics[scale=1]{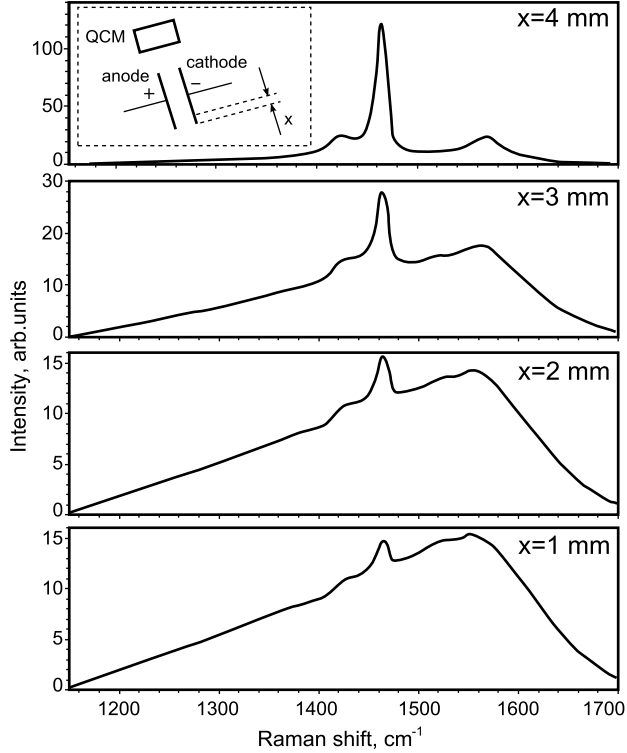}
		\caption{Raman spectra of the cathode coating deposited using scheme 2 (inset) taken at various distances $x$ from the front edge of the electrode.}
\end{center} 
\end{figure}

Taking into consideration the approximate width of the strongly modified coating area formed near the front edge of the cathode (3~mm) and assuming that structural modification was induced by \ce{C60+} ions, the kinetic energy of the majority of positive fullerene ions produced by fullerite EBD is estimated to be up to 7~eV. At the same time, the kinetic energy of fullerene molecules estimated from the temperature of the target surface ($T=673$--1000 K~\cite{20}) is less than 1~eV (0.087--0.13~eV). This inconsistence shows that positive charging of the fullerite target up to several volts occurs under electron-beam irradiation. Furthermore, such charging was also confirmed by the previously experimentally observed scattering of the fullerite powder during EBD.\\

In deposition schemes 3 and 4, the substrates were placed onto the electrode behind the grounded copper~mesh. In comparison with schemes 1 and 2, schemes 3 and 4 allow the deposition of uniform layers and, at the same time, the electrostatic field does not affect fullerite EBD. The influence of positive fullerene ions on the growing coating structure was studied by applying electrostatic potential from +10~V to -300~V to the electrode. Thus, \ce{C60} films were deposited under the conditions ranging from the absence of positive ions to bombardment by the accelerated positive ion flux. Scheme 4 has the smaller electrode size, inducing additional increase of the charged particle flux as a result of electrostatic focusing. \\

Raman spectra of the coatings deposited using schemes 3 and 4 were decomposed, by Lorentzian line-shape analysis, into the broad continuum component and peaks of fullerene monomers, dimers, linear chains and 2D polymer networks, similarly to the spectrum in Fig. 2b. The broad continuum in the range of 1100--1700~cm$^{-1}$ appears for the layers formed at -100~V and dominates the spectrum for higher negative potential. Raman bands characteristic of \ce{C60} polymers dominate the spectra of the layers deposited at the substrate potentials from +10~V to -50~V and become less intensive for higher negative potential. Relative contents of different \ce{C60} polymers and the highly cross-linked carbon phase are estimated by dividing the area of the corresponding A$_g$(2) mode or the area of the broad continuum by the area of the \ce{C60} monomer A$_g$(2) mode for each spectrum (Fig. 4). \\

The relative content of \ce{C60} polymer phases decreases with increasing negative potential until the value of 50-100 V. We attribute this result to the decrease in the number of low-energy secondary electrons being able to reach the substrate. The higher the substrate negative potential, the less the secondary electron bombardment that occurs. Whereas the layer modification induced by the positive fullerene ions accelerated up to 50--100~eV still plays a minor role, the lack of secondary electrons decreases the coating polymerization. With further increase of the negative potential, the role of positive fullerene ions bombardment becomes more significant, the relative content of \ce{C60} polymer phases increases, and the broad continuum appears in the Raman spectrum. Electrostatic focusing of positive fullerene ion flux onto the substrate by using scheme 4 results mainly in a considerable increase of 2D and broad continuum components (Figs. 4c,d) at potentials from -100 to -300 V, which suggests higher cross-linking of the deposited layer.\\

\begin{figure}[h!]
 \begin{center}
 		\includegraphics[width=\textwidth ]{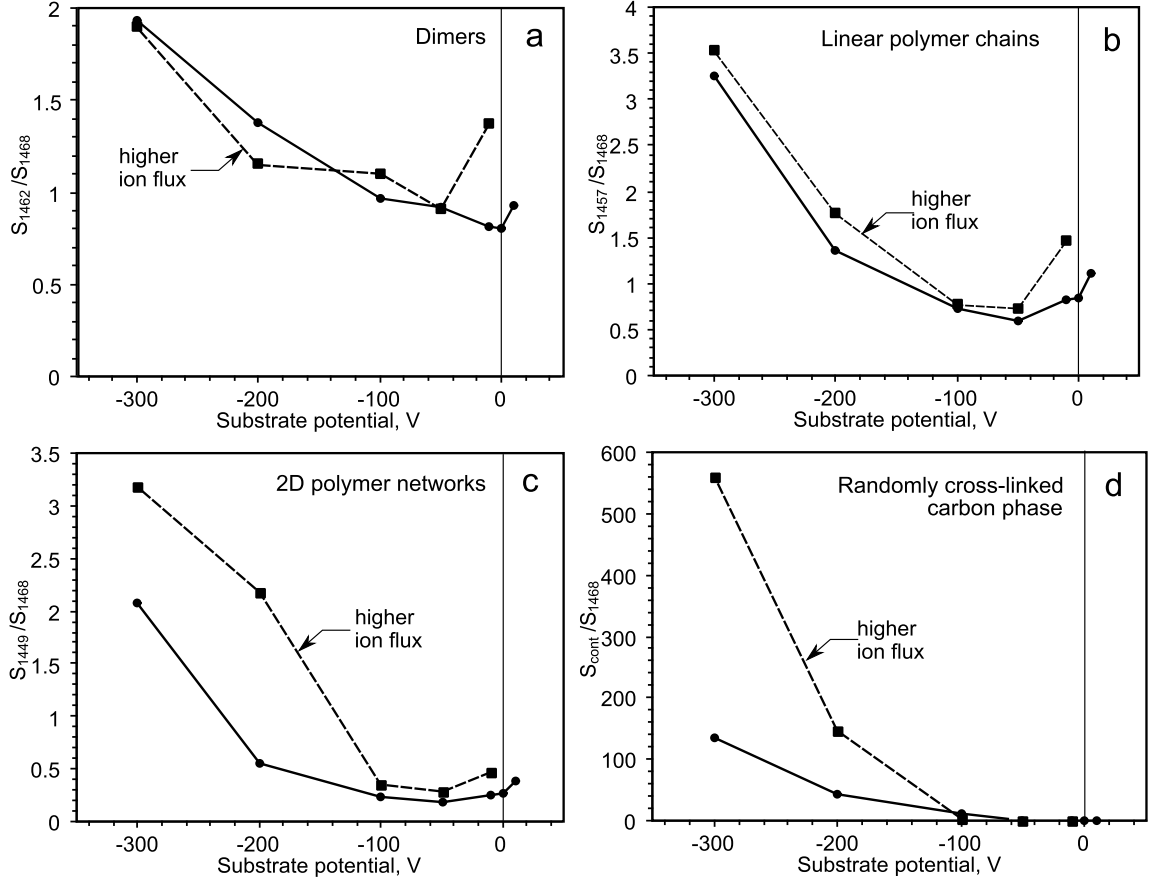}
		\caption{Ratio of the A$_g$(2) peak area of \ce{C60} polymers and the area of the broad continuum to the area of the A$_g$(2) peak of \ce{C60} monomers at 1468~cm$^{-1}$: dimers (a), linear chains (b), 2D polymer networks (c), broad continuum (d). Solid and dashed lines represent the Raman spectra of the coatings deposited using scheme 3 and scheme 4, respectively.}
\end{center} 
\end{figure}

Four strong IR absorption bands with F$_{1u}$ symmetry at 526, 576, 1184 and 1428~cm$^{-1}$, characteristic of pristine fullerene \ce{C60}, decrease and practically disappear from the spectra of the deposited coatings with increasing negative substrate potential, indicating the decrease of the fullerene monomer content (Fig. 5). The absorption band at 526~cm$^{-1}$ considerably decreases but still remains in the spectra. However, this may be due to the appearance of new bands of dimers and linear polymer chains in close proximity~\cite{10}. Additional bands assigned to fullerene polymers appear in the spectra of all deposited layers. For example, linear chains and 2D polymer networks contribute to the absorption at around 1260~cm$^{-1}$. The doublet at 1342/1346~cm$^{-1}$ is characteristic of 2D fullerene polymer networks. Compared with the 526~cm$^{-1}$ band, the area of the 1342/1346~cm$^{-1}$ doublet decreases up to the negative substrate potential of 50-100~V and then increases for higher negative potentials (the inset in Fig. 5). This increase at potentials of -200 and -300~V is more intensive for concentrated positive ion flux (layers deposited using scheme 4) and corresponds well with the Raman spectroscopy results. On the other hand, some new broad absorption bands not typical of fullerene dumb-bell-shaped polymers appear in the FTIR spectra.\\

We attribute the strong broad peak at around 1400~cm$^{-1}$ to peanut-shaped fullerene polymers. A similar absorption band was previously reported by Onoe and Takeuchi to be a dominant peak for thin fullerene films irradiated by a 3~keV electron beam and it was assigned to the vibration of the waist region in peanut-shaped structures~\cite{18}. An absorption band at 874~cm$^{-1}$ was previously observed in IR spectra of single-wall carbon nanotubes~\cite{22} and thus, it also can be connected to the formation of peanut-shaped linear fullerene polymer chains. A broad peak at around 1100~cm$^{-1}$ was previously reported by Rao \textit{et al.}~\cite{9} for HPHT fullerene polymers. It was attributed to disordered or incomplete covalent bonds between \ce{C60} molecules. However, if such incomplete carbon bonds are formed during coating growth in vacuum, they can react with oxygen after exposure to the atmosphere. It is well known that fullerite easily absorbs oxygen molecules into its interstitial sites, and under normal conditions, about 1\% of them is filled with oxygen. After exposure to light, even commercially available pure as-produced fullerite, by the moment of practical usage, already contains some percent of \ce{C120O} dimer that cannot be eliminated by~even vacuum degassing and that reacts further, resulting in oxygen-containing polymer formation~\cite{23}. Moreover, absorption at around 1100~cm$^{-1}$ and 960~cm$^{-1}$ is characteristic for epoxide derivatives and therefore, we attribute these broad bands to fullerene oxides.\\

\begin{figure}[h!]
 \begin{center}
 		\includegraphics[scale=1]{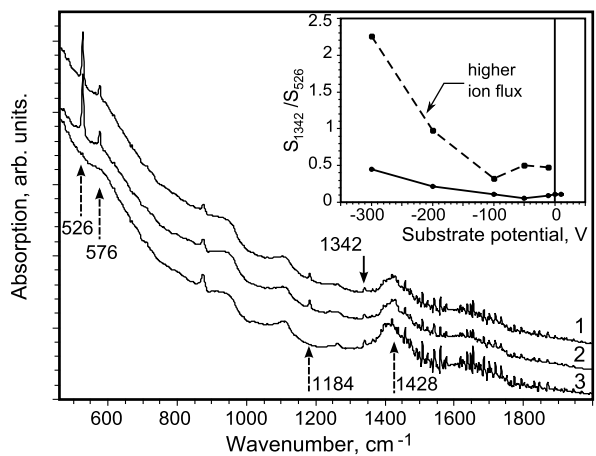}
		\caption{FTIR spectra of \ce{C60} coatings deposited at 0~V (1), +10~V (2) and -300~V (3) substrate potentials. Dashed arrows show the position of the bands of pristine fullerite \ce{C60}. Solid arrow shows the position of the doublet at 1342/1346~cm$^{-1}$. The inset shows the ratio of the 1342/1346~cm$^{-1}$ doublet area to the area of the 526~cm$^{-1}$ band for the coatings deposited at different substrate potentials using scheme 3 (solid line) and scheme 4 (dashed line).}
\end{center} 
\end{figure}

Upon fullerene film polymerization and cross-linking, covalent bonding between \ce{C60} molecules leads to the transformation of the carbon atom local coordination from sp$^2$ to sp$^3$-like. The XPS C 1s core peak is a good indicator of such a structural transformation, as it shifts from 284.4~eV for sp$^2$ in graphite to 285.2~eV for sp$^3$ in diamond. In our case, the binding energy of the C 1s core peak for the coating deposited at the -300~V substrate potential was approximately 0.2~eV higher than that for the coating deposited without electrostatic potential (0 V). The full width at half-maximum (FWHM) of the C 1s peak also increased by \textit{ca.} 0.2~eV toward the higher binding energy region (Fig. 6). This corresponds well with the increase of the sp$^3$ coordinated carbon site content proposed on the basis of the Raman spectroscopy results (higher cross-linking of the layers). It also should be noted that the observed C 1s peak shift and broadening are more intensive than those for photopolymerized \ce{C60} films. Onoe \textit{et al.} reported only a slight C 1s line shift and peak broadening of less than 0.1~eV after 200 hours of fullerene film UV-Vis irradiation in vacuum when the average number of four-membered ring covalent links per \ce{C60} molecule in the layer was already about four~\cite{24}. Thus the obtained coatings indeed significantly differ from photopolymerized layers. However, let us next consider the question of whether the fullerene molecules are still characteristic structural elements of these layers or the layers rather represent disordered carbon coatings similar to diamond-like carbon films.\\

\begin{figure}[h!]
 \begin{center}
 		\includegraphics[scale=1]{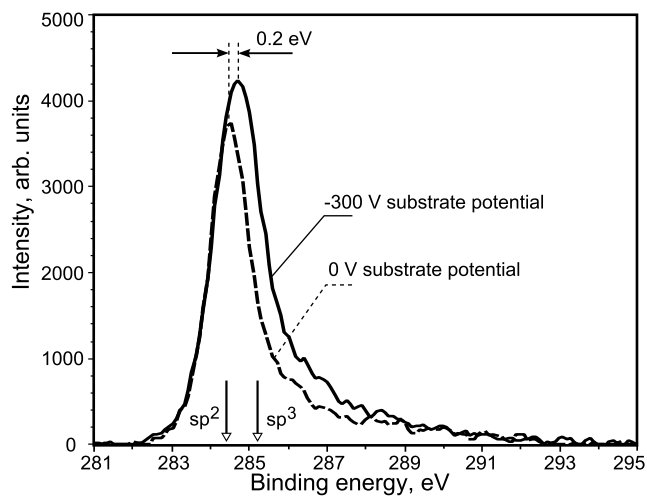}
		\caption{XPS spectra of C 1s core peak of fullerene film deposited at -300~V (solid line) and 0~V (dashed line) substrate potentials.}
\end{center} 
\end{figure}

LDI mass-spectrometry measurement was performed on the films deposited using schemes 3 and 4. Before starting the analysis of the obtained mass spectra it should be noted that dumb-bell-shaped \ce{C60} polymers dissociate upon LDI. Wang \textit{et al.}~\cite{25} showed, using several mass-spectroscopy techniques, that the spectra of the highly purified dumb-bell-shaped \ce{C120} dimer in both negative and positive ions show a strong peak of the \ce{C60} monomer, and only a weak series of mass peaks corresponding to the dimer and its fragments was detected using Fourier-transform ion cyclotron resonance (FTICR) and matrix-assisted laser desorption ionization (MALDI) techniques. This means that dumb-bell-shaped dimers and higher polymer chains/clusters easily dissociate to monomers or smaller polymer chains/clusters during desorption and ionization. On the other hand, Hara \textit{et al.} showed that peanut-shaped fullerene polymers are more stable during LDI and hardly dissociate to monomers during the measurement~\cite{17}. Therefore we should remember that the obtained mass spectra are strongly influenced by complex physical chemical processes in the material induced by the laser irradiation during the measurement. Nevertheless, LDI mass spectrometry allows us to distinguish between disordered carbon and randomly cross-linked fullerene molecules.\\

\begin{figure}[h!]
 \begin{center}
 		\includegraphics[scale=1]{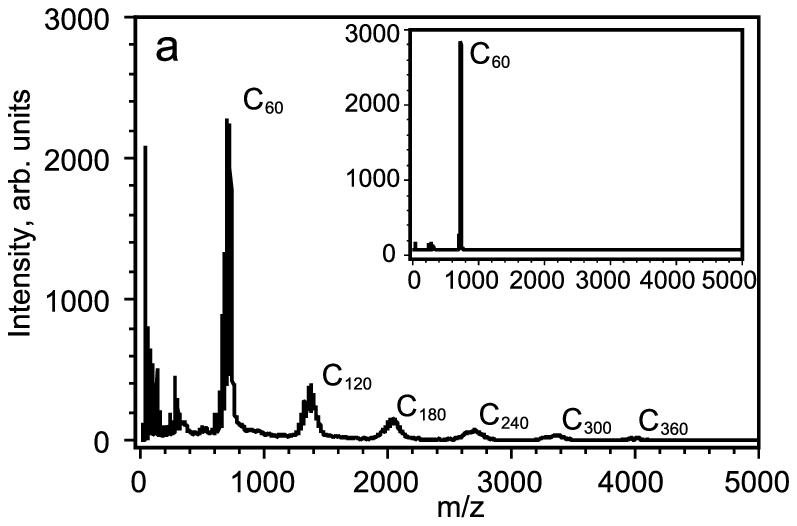}
		\includegraphics[scale=1]{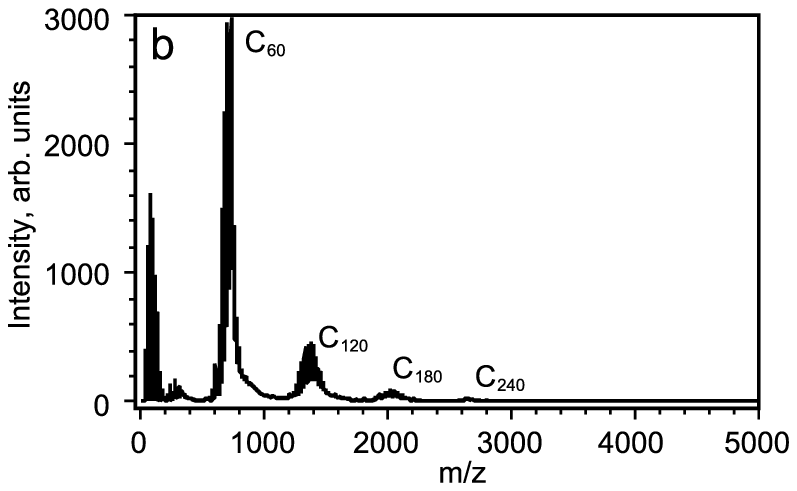}		
		\caption{LDI mass spectra of negative ions of the pristine fullerite powder (inset), the layer deposited without electric potential on the substrate (a) and the layer deposited at the constant substrate potential of -300~V (b).}
\end{center} 
\end{figure}

The negative ion mass spectrum of the pristine fullerite (Fig.~7a) consists of a strong peak of \ce{C60} ($m/z=720$), considerably less intensive peaks of \ce{C60O} ($m/z=736$), \ce{C58} ($m/z=696$), and some low molecular fragments. No \ce{C120} or higher \ce{C60} oligomers were detected, which indicates the laser power used induces negligible fullerene polymerization. The mass spectra of both the coating deposited without applying substrate potential (Fig.~7b) and the coating deposited at the highest studied negative substrate potential of -300~V (Fig.~7c) are dominated by the group of peaks at around $m/z=720$ representing \ce{C60} and its derivatives formed by \ce{C2} loss or gain. It is unlikely for fullerene molecules to be formed in such a quantity from the disordered carbon fragments during LDI. Furthermore, the content of low molecular fragments in the mass spectra of the deposited coatings amounts to a few mass percent and changes insignificantly with increasing negative substrate potential. This means that for all studied coatings, fullerene molecules are still characteristic structural elements. Thus,~even the coating deposited under the bombardment of positive fullerene ions with the energy of 300~eV represents a randomly cross-linked disordered 3D fullerene network rather than the disordered carbon layer. Both mass spectra in Fig.7 show groups of peaks of fullerene dimers and higher oligomers. Though the mass spectrum of the coating deposited without applying electrostatic potential shows mass peaks of larger fullerene oligomers (up to \ce{C360}) than the mass spectrum of the coating deposited at the substrate potential of -300~V (up to \ce{C240}), it can indicate that the latter coating is more difficult to evaporate and ionize because of its higher degree of cross-linking.\\

\begin{figure}[h!]
 \begin{center}
 		\includegraphics[scale=1]{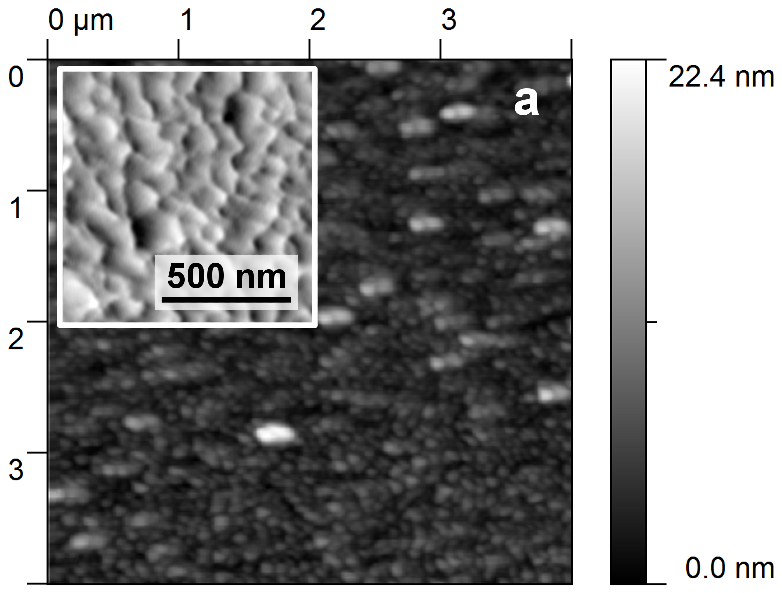}
		\includegraphics[scale=1]{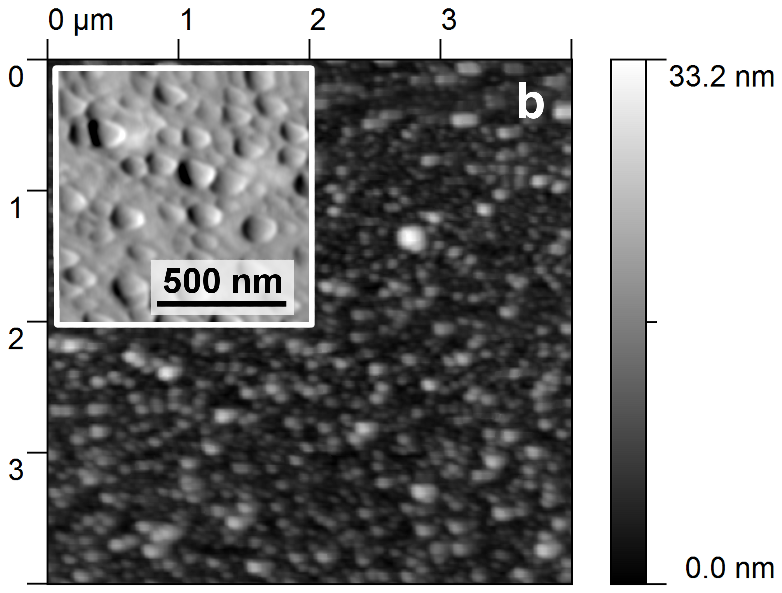}		
		\caption{AFM topography images of the layers deposited at 0~V (a) and -300~V (b) substrate potential. Insets show magnified and shaded topography images.}
\end{center} 
\end{figure}

Fullerene polymerization induces material hardening and, as was already mentioned, 3D polymerized samples exhibit ultrahigh mechanical properties. In the case of multiphase coatings containing different fullerene polymer structures, mechanical properties of the layer should also exhibit heterogeneity across the surface. However, the phase contrast images obtained by AFM in the dynamic force mode for all deposited layers are highly uniform, which means that at the level of AFM resolution used, the mechanical properties of the coatings appear homogeneous across the surface. Despite the significant difference in molecular structure, similar surface clusters of 50--400~nm lateral size can be observed in the AFM topography images of all studied coatings (Fig. 8). The coatings deposited at higher negative substrate potentials exhibit slightly more weakly structured surface areas. No pores were found by AFM measurement. All the coatings exhibit low roughness with the $R_a$ parameter (arithmetic average of absolute height values) increasing approximately from 1 to 3~nm with an increase of the negative substrate potential. For the pristine substrate (Si monocrystal with 100-nm-thick sublayer of Al), the $R_a$ parameter was about 0.5~nm. The low phase contrast at the resolution level used and the uniform morphology of the coatings suggest that the deposited layers consist of a highly dispersed and homogeneous mixture of the various types of fullerene polymer structures detected by the spectroscopic measurements.

\section{Summary}
The charge composition of fullerite powder EBD products and the influence of the fullerene ions and electrons bombardment on the fullerene film growth and structural modification were studied by Raman, ATR FTIR, XPS, LDI MS and AFM measurements of the layers deposited with additional electric field assistance. The main results can be summarized as follows.\\

It was found that fullerite powder EBD products contain electrons and positive fullerene ions (a few mass percent of the evaporating molecular flow under the EBD conditions used). The energy of the positive fullerene ions was estimated to be approximately up to 7~eV, and thus positive charging of the fullerite target occurs during the EBD process. Bombardment of the growing layer by electrons additionally accelerated up to 300~eV by the substrate potential results in the formation of polymerized \ce{C60} coatings containing dimers, linear chains and 2D networks of dumb-bell-shaped \ce{C60} polymers as well as peanut-shaped \ce{C60} polymers, similarly to the coatings deposited without electric field assistance. Bombardment of the growing \ce{C60} layer by positive fullerene ions accelerated to the energy of up to 300~eV leads to the formation of a highly cross-linked random 3D network of covalently bonded \ce{C60} molecules.
\section*{Acknowledgements}
This work was financially supported by the Shizuoka University, The True-Nano Project and the Belarusian Republican Foundation for Fundamental Research.
\bibliographystyle{ieeetr}
\bibliography{razanau_TSF}
\end{document}